\documentclass[aps,prd,twocolumn,groupedaddress,nofootinbib,showpacs, showkeys,amssymb,10pt]{revtex4}
\usepackage{latexsym,float}
\usepackage{epsfig,graphics}
\usepackage{epstopdf}
\usepackage{amssymb}
\usepackage{amsmath}
\usepackage{verbatim}
\usepackage{multirow}
\usepackage{color}
\usepackage{tikz}
\usepackage[utf8]{inputenc}
\usetikzlibrary{matrix}

\def\gtwid{\mathrel{\raise.3ex\hbox{$>$\kern-.75em\lower1ex\hbox{$\sim$}}}}
\def\ltwid{\mathrel{\raise.3ex\hbox{$<$\kern-.75em\lower1ex\hbox{$\sim$}}}}
\def\square{\kern1pt\vbox{\hrule height 1.2pt\hbox{\vrule width 1.2pt\hskip 3pt
			\vbox{\vskip 6pt}\hskip 3pt\vrule width 0.6pt}\hrule height 0.6pt}\kern1pt}
\begin{document}

\title{Phase Space description of  Nonlocal Teleparallel Gravity}

\author{Kazuharu Bamba}
\email{bamba@sss.fukushima-u.ac.jp}
\affiliation{ Division of Human Support System,\\
Faculty of Symbiotic Systems Science,\\
Fukushima University, Fukushima 960-1296, Japan}

\author{Davood  Momeni}
\email{davood@squ.edu.om}
\affiliation{Department of Physics, College of Science,\\ Sultan Qaboos University,
\\P.O. Box 36, P.C. 123, Al-Khodh, Muscat, Sultanate of Oman}
\author{Mudhahir Al Ajmi}
\email{mudhahir@squ.edu.om}
\affiliation{Department of Physics, College of Science,\\ Sultan Qaboos University,
\\P.O. Box 36, P.C. 123, Al-Khodh, Muscat, Sultanate of Oman}

\begin{abstract}
We study cosmological solutions in nonlocal teleparallel gravity or $f(T)$ theory, where 
$T$ is the torsion scalar in teleparallel gravity. This is a natural extenstion of the usual teleparallel gravity with nonlocal terms. In this work the phase space portrait proposed to describe the dynamics of an arbitrary flat, homogeneous cosmological background with a number of matter contents, both in early and late time epochs. The aim was to convert the system of the equations of the motion to a first order autonomous dynamical system and to find fixed points and attractors using numerical codes. For this purpose,  firstly we derive effective forms of cosmological field equations describing the whole cosmic evolution history 
in a homogeneous and isotropic cosmological background and construct the autonomous system of the first order dynamical equations. 
In addition, we investigate the local stability in the dynamical systems called ``the stable/unstable manifold'' by introducing a specific form of the interaction between matter, dark energy, radiation and a scalar field. 
Furthermore, we explore the exact solutions of the cosmological equations in the case of de Sitter spacetime. In particular, we examine the role of an auxiliary function called ``gauge'' $\eta$ in the formation of such cosmological solutions and show whether the de Sitter solutions can exist or not. 
Moreover, we study the stability issue of the de Sitter solutions both in vacuum and non-vacuum spacetimes. It is demonstrated that for nonlocal $f(T)$ gravity, the stable de Sitter solutions can be produced even in vacuum spacetime. 
\end{abstract}
\keywords{Teleparallel gravity; phase space analysis; attractors; dynamical systems }
\date{\today}

\maketitle

\section{Introduction} \label{introduction}

It has been supported that 
in addition to the inflationary stage~\cite{Inflation} in the early universe, 
currently the expansion of the universe is also accelerating 
by various cosmological observations including 
Type Ia Supernovae~\cite{SN}, 
cosmic microwave background (CMB) radiation~\cite{Ade:2015xua}, 
large scale structure~\cite{LSS}, 
baryon acoustic oscillations (BAO)~\cite{Eisenstein:2005su} 
as well as weak lensing~\cite{Jain:2003tba}. 
We have two representative explanations for such a 
late-time cosmic acceleration. 
One approach is to introduce ``dark energy'' (DE) in the context of 
general relativity. 
The other approach is to consider the modification of gravity on the large scale (for reviews on not only DE problem but also 
modified gravity theories, see, for example,~\cite{R-DE-MG}). 

There is a possible candidate for a theory of gravitation alternative to general relativity, namely, teleparallel gravity, which is described by 
using the Weitzenb\"{o}ck connection~\cite{T-G}. 
In teleparallel gravity, there exists torsion. 
This is opposite to the case of general relativity, in which the Levi-Civita connection is used. 
The torsion scalar $T$ represents the Lagrangian density 
of teleparallel gravity. 
It can be extended to a function of $T$, that is, $f(T)$ gravity (for a recent review, see, for instance,~\cite{Cai:2015emx}). 
This idea is similar to that of $f(R)$ gravity~\cite{F-R}, where 
$R$ is the scalar curvature. 
Inflation in the early universe~\cite{F-T-Inf} and the late-time cosmic acceleration~\cite{F-T-LC} can be realized 
in $f(T)$ gravity. 
Various cosmological and astrophysical considerations in $f(T)$ gravity have widely been executed~\cite{F(T)-Refs}. 
It is known that in $f(T)$ gravity, the local Lorentz invariance is broken~\cite{L-L-I}, and the relevant investigations on this point have been discussed~\cite{RP-LLI}. 

On the other hand, in Ref.~\cite{Deser:2007jk}, 
there has been considered a way of modifying gravitation, 
the so-called nonlocal gravity, which comes from quantum effects. 
Furthermore, in order to unify inflation in the early universe 
and the late-time accelerated expansion of the universe, 
non-local gravity has been modified by adding an $f(R)$ term 
in Ref.~\cite{Nojiri:2007uq}. 
In addition, a possible solution for the cosmological constant problem 
through the nonlocal property of gravitation~\cite{ArkaniHamed:2002fu} 
has been proposed. 
Moreover, a physical mechanism by which a cosmological constant is screened 
in the framework of nonlocal gravity has been investigated~\cite{Nojiri:2010pw, Bamba:2012ky, Zhang:2011uv}. 
It has also been indicated that in nonlocal gravity, 
there is the issue of ghosts~\cite{Nojiri:2010pw}. 
Various aspects of nonlocal gravity have widely been explored~\cite{NL-Ref} (for a recent review on nonlocal gravity, see, e.g.~\cite{Maggiore:2016gpx}). 
It is worth noting that nonlocal terms $\Box T $ was first used in the framework of modified teleparallel gravity in Ref.\cite{Otalora:2016dxe}
Furthermore , the nonlocal deformations of teleparallel gravity 
have been analyzed in Refs.~\cite{Bahamonde:2017bps},\cite{Channuie:2017txg}. 
This theory is called nonlocal $f(T)$ gravity, which 
can be considered as an extension of nonlocal general relativity to the Weitzenb\"{o}ck spacetime. 
It has been discussed that there is a possibility to distinguish teleparallel gravity from general relativity by future experiments detecting nonlocal effects.  
In this paper, we investigate exact cosmological solutions in nonlocal $f(T)$ gravity. We analyze the autonomous system of the first order dynamical equations 
by deriving effective forms of cosmological field equations in a homogeneous and isotropic cosmological background, describing the whole evolution history of the universe. 
Moreover, we propose a specific form of the interaction between matter, dark energy, radiation and a scalar field and examine the local stability in the dynamical systems, which is called 
``the stable/unstable manifold''. As a result, it is demonstrated that the system has a stable attractor. 
Furthermore, we study exact solutions of the cosmological equations in the case of de Sitter spacetime. Particularly, we explore the role of an auxiliary function called ``gauge'' $\eta$ in the formation of such cosmological solutions 
and show whether the de Sitter solutions can exist or not in this scenario. 
In addition, we consider the stability problem of the de Sitter solutions both in vacuum and non-vacuum spacetimes and find that even in vacuum spacetime, the stable de Sitter solutions can be produced in the framework of nonlocal $f(T)$ gravity.

The organization of the paper is the following. 
In Sec.~2, we explain the framework of nonlocal $f(T)$ gravity. 
In Sec.~3, we explore the cosmological background and effective field equations in nonlocal $f(T)$ gravity. 
In Sec.~4, the interaction term and phase portrait are analyzed. 
In Sec.~5, the de Sitter solution is derived and its stability is 
examined in the following Sec.~6. 
Finally, conclusions are provided in Sec.~7.  
\section{Formal framework of nonlocal $f(T)$ gravity}
Let us develop the formalism of nonlocal modified gravity with torsion $T$ in a same manner as the nonlocal $f(R)$ gravity is developed \cite{Nojiri:2007uq}. We suppose that the possible action for gravity with matter contents is given in terms of classical gauge invariant action as follows:
\begin{eqnarray}
&&S=\frac{1}{2\kappa}\int d^4x eT\Big(f(\Box^{-1}T)-1\Big)+\int d^4x e \mathcal{L}_m\label{action}
\end{eqnarray}
where $\kappa=8\pi G$, $G$ is Newtonian gravitational constant, $\mathcal{L}_m$ is matter Lagrangian. To describe the geometry of spacetime in teleparallel gravity, it is commonly used the tetrads formalism where the metric can be written in an orthogonal frame $e_a^{\mu}$, in a such manner that $g^{\mu\nu}=e_{a}^{\mu}e_b^{\nu}\eta^{ab}$, where Greek alphabets run from $\mu,\nu=0...3$, the flat Minkowski metric is denoted by $\eta^{ab}$. Note that $e_a^{\mu}e^{a}_{\nu}=\delta_{\nu}^{\mu}$ and $\Box^{-1}$ is considered as an integral over the entirely spacetime manifold. The local  operator  $\Box =\nabla^{\mu}\nabla_{\mu}$ is called  d'Alembert operator defined as  $\Box=e^{-1}\partial_
{\alpha}(e\partial^{\alpha})$ here $e=det(e_{a}^{\mu})=\sqrt{-det(g_{\alpha\beta})}$ and $T$ is torsion scalar and it is defined in a same form as $f(T)$ gravity.

It is always possible to reduce nonlocal theories to scalar-tensor equivalent theories  and it is easy to do that for our model given in (\ref{action}) using two auxiliary (non ghost) fields $\phi=\frac{1}{\Box}T$ and $\xi=-\frac{1}{\Box}(f'(\phi)T)$. The reason that those fields are considered as non ghost is that , the norm of them defined as $||\phi||=\int_{\Sigma}|\phi|^2 ed^4x$ is always positive definite and never becomes complex as long as the metric and its torsion $T$ remains real numbers. As long as we work in Riemanninan manifolds this condition will be hold and we can safely use them as an appropriate set of auxiliary fields. 
\par
The new form for the reduced action is written as follows:
\begin{eqnarray}
&&S=\frac{1}{2\kappa}\int d^4x e\Big[T\Big(f(\phi)-1\Big)-\partial_{\mu}\xi\partial^{\mu}\phi-\xi T
\Big]\\&&\nonumber
+\int d^4x e \mathcal{L}_m\label{action2}.
\end{eqnarray}
Note that in (\ref{action2}) the action function $f(\phi)$ is supposed to have any desired form.  Formally if we take the case: $\xi =1$ and $f(\phi) = 2$ in action of theory given by Eq. (2), then the action is reduced to the one which is equivalently of Teleparallel Gravity for $T\neq 0$. Classical tests for GR prove a very good agreement with observations. As a result it is very important to know whether this nonolocal teleparallel gravity has GR limit or not. At the level of action we already demonstrate it. By an enough good choosing of the function $f(\phi)$ we can recover GR as a limiting case.

\par
 The form of equations of motion is presented in  \cite{Bahamonde:2017bps} :

\begin{eqnarray}
&&2(1-f(\phi)+\xi)\left[ e^{-1}\partial_\mu (e S_{a}{}^{\mu\beta})-E_{a}^{\lambda}T^{\rho}{}_{\mu\lambda}S_{\rho}{}^{\beta\mu}-\frac{1}{4}E^{\beta}_{a}T\right]\\&&\nonumber
-\frac{1}{2}\Big[(\partial^{\lambda}\xi)(\partial_{\lambda}\phi)E_{a}^{\beta}-(\partial^{\beta}\xi)(\partial_{a}\phi)-(\partial_{a}\xi)(\partial^{\beta}\phi)\Big] \\&&\nonumber-2\partial_{\mu}(\xi-f(\phi))E^\rho_a S_{\rho}{}^{\mu\nu}=  \kappa\Theta^\beta_a\,. \label{2}\\&&
\Box\xi+Tf'(\phi)=0,\\&&
\Box\phi-T=0.
\end{eqnarray}

here the tensor $E_{\alpha}^{\beta}$ is defined through the variation of $e$ as follows
$\delta e=e E^{\beta}_\alpha e_{\beta}^a$, $\Theta^\beta_a$  is the energy-momentum tensor of matter contents defined by $\Theta^\beta_a=e^{-1}\frac{\delta (e \mathcal{L}_m)}{\delta e_{\beta}^a}$ and $\Box\equiv e^{-1}\partial_{\mu}(e\partial^{\mu}) $. 

In Refs. \cite{Bahamonde:2017bps} , the authors investigated cosmological data analysis on a suitable chosen function $f(\phi)=A \exp(n\phi)$ and later in Ref. \cite{Channuie:2017txg} , using Noether symmetry approach. In our paper we will fix $f(\phi)$ in another simple/adequate form in next section.

\section{Cosmological background and effective field equations}\label{fieldeqs}
%
The aim of this section is to write equations of motion for a cosmological background in the presence of matter fields in an effective form. Let us  suppose that the non singular, physical metric of spacetime is given in the form of a  Friedman-Lemaitre-Robertson-Walker (FLRW) metric  given by $ds^2=dt^2-a(t)^2(dx^bdx_b)$, where $b=1,2,3$ is spatial coordinate and $a(t) $ is scale factor and measures expansion of the whole cosmological Universe as well as its acceleration/deceleration phase. The corresponding suitable, diagonal tetrads basis is given by $e^{a}_{\mu}=diag\Big(1,a(t),a(t),a(t)\Big)$. The set of FLRW equations and the equations for the scalar fields are written as follow:
\begin{widetext}
\begin{eqnarray}
&&3H^2(1+\xi-f(\phi))=\frac{1}{2}\dot{\phi}\dot{\xi}+\kappa(\rho_m+\rho_{\Lambda}+\rho_r)\label{eq1}\\
&&(2\dot{H}+3H^2)(1+\xi-f(\phi))=-\frac{1}{2}\dot{\phi}\dot{\xi}+2H(\dot{\xi}-\dot{f}(\phi))-\kappa(p_{\Lambda}+p_r)\label{eq2}\\&&
\ddot{\xi}+3H\dot{\xi}-6H^2f'(\phi)=0
\label{eq3}
\\&&\ddot{\phi}+3H\dot{\phi}+6H^2=0
\label{eq4}
\end{eqnarray}
\end{widetext}
The matter energy-momentum tensor is given in terms of a diagonal tensors for matter , dark energy, radiation as follows:
\begin{widetext}
\begin{eqnarray}
&&\tau_{\mu}^{\nu}=e^{a}_{\mu}e_{b}^{\nu}\tau_{a}^{b}=diag\Big(\rho
_m+\rho_{\Lambda}+
\rho_r,-p_{\Lambda}-
p_r,-p_{\Lambda}-
p_r,-p_{\Lambda}-
p_r\Big)
\end{eqnarray}
\end{widetext}
where $e^{a}_{\mu}e^{\mu}_{a}=\delta^{a}_{b}$ is unit matrix.  The matter budget of our model is drak matter density $\rho_m$,  radiation field density $\rho_r$ and scalar field density $\rho_{\phi}$. In order to preserve the acceleration expansion and the existence of late time de Sitter cosmology we inserted a non zero cosmological constant $\Lambda$ with energy density $\rho_{\Lambda}$. As an attempt to keep simplicity 
we assume that all matter contents are given in barotropic forms, where we define the equation of state (EoS) parameter $w_a$ for each fluid component, namely matter, radiation and cosmological constant and as a result for any component of matter field we have a linear EoS , i.e, $p_a=w_a\rho_a$. Here the roman index $a$ refers to different matter contents. Namely we denote it by $a =\{m,\Lambda,r\}$ where $m$ is for matter, $\Lambda$ is for DE and $r$ is for radiation field. Note that neither $\phi$ nor $\xi$ are considered as the DE. The reason is that both fields play the role of auxiliary fields. We can't make guarantee that whether  the fields $\phi,\xi$  will be ghost or not. Actually the appearance of ghost scalar fields in the non local theories for gravity is an important issue and should be addressed adequately. For example in the non local extensions of the GR, when the action is corrected by nonlocal terms $\frac{1}{\Box}R$ or higher order terms, one must count the number of the degrees of freedom of the localized form of the Lagrangian. Additionally, one needs to check the equivalence between local and nonlocal representations of theory both at the action level and equations of motion levels. It is possible to make a categorization based on the first form of the auxiliary fields. Based on this classification we can find the number of algebraic constraints which they will limit our ability to write the local or nonlocal representations of theory. Although in nonlocal extensions of the GR, as long as we have a linear term we can ascertain the equivalence between frames. However, with higher order terms this equivalence is broken. That means in a general nonlocal GR when we have only curvature terms our theory may suffer from ghosts.
In nonlocal extensions of the teleparallel gravity (TEGR), we can deduce the same as long as we make the theory using a 
nonlocal action made by linear scalar torsion $T$, and, hence, no ghost will appear. That is because it was proved that the Einstein-Hilbert action GR is dynamically equivalent to the TEGR at level of action as well as the equations of motion \cite{Hayashi}. In our study with the nonlocal term which we will opt in next paragraph the model will be ghost free. Although probably the scalar field $\phi$ will not be a ghost but still we do not have any strong reason to keep it as the only sector for the acceleration expansion in our model. For this reason we also keep the cosmological constant $\Lambda$ and its energy density $\rho_{\Lambda}$.

The first challenge is to choose one suitable form for $f(\phi)$.  Note that $\Box\xi=-f'(\phi)\Box\phi$. It is illustrative to expand and write this equation in the following equivalent form : 
$$\Box(\xi+f(\phi))=f''(\phi)\Box\phi$$
Here, we suppose that $f''(\phi)=0$. Thus one suitable class of models is:
\begin{eqnarray}
&&f(\phi)=A\phi+B\label{fphi}.
\end{eqnarray}
Note that for $A=0$ , $B=2$ the results reduce to the GR as a limiting case.
Note that now, $\Box(\xi+f(\phi))=0$, and we have a freedom to take $\xi+f(\phi)=\Psi$, where $\Psi$ is a harmonic function over $\mathcal{R}^4$. A possible option is to consider  $\Psi=2A\phi$, consequently the set of eqs. (\ref{eq1},\ref{eq2}) are simply written in the following forms:
\begin{eqnarray}
&&3H^2=\frac{\frac{1}{2}A\dot{\phi}^2+\kappa(\rho_m+\rho_{\Lambda}+\rho_r)}{1-2B}\label{eq11}\\
&&2\dot{H}+3H^2=-\frac{\frac{1}{2}A\dot{\phi}^2+\kappa(p_{\Lambda}+p_r)}{1-2B}\label{eq22}
\\&&
\ddot{\xi}+3H\dot{\xi}-6AH^2=0
\label{eq33}
\\&&\ddot{\phi}+3H\dot{\phi}+6H^2=0
\label{eq44}
\end{eqnarray}
Here parameter $B$ measures the difference between TEGR and nonlocal theory respectively. Because we will study time evolution of the energy densities, it is adequate to rewrite cosmological equation , presented in Eq. (\ref{eq11}) in the following forms, 
\begin{eqnarray}
&& 3H^2=\frac{\rho_{\phi}}{1-2B}+\frac{\rho_{m}}{1-2B}+\frac{\rho_{\Lambda}}{1-2B}+\frac{\rho_{r}}{1-2B}\label{eq111}
\end{eqnarray}
in the above equation, we have defined
\begin{eqnarray}
&& \rho_{\phi}\equiv A\dot{\phi}^2
\end{eqnarray}
Note that the other density functions can not be written explicitly in terms of the scale factor $a$ or scalar fields, till the time which we will present continuity equations for all the matter components. In our scenario we assumed that different matter contents interact with each other through some interaction forms which will present in next section. 

\subsection{Hartman-Grobman linearizion theorem}\label{Hartman-Grobman}
To investigate the phase space analysis it is needed to reduce the system of equations to an autonomous system of first order differential equations in the form $\frac{d\vec{X}}{dN}=f(\vec{X})$, where $N$ plays the role of time and $\vec{X}$ is a vector field with density functions as components.In many works, investigation of various aspects of dynamical systems in cosmology of modified gravity is discussed \cite{Odintsov:2015wwp}.
\par
The Hartman-Grobman linearizion theorem provides a powerful technique to study the local stability and the portrait of the phase space, when we have a set of hyperbolic fixed points . Let $\vec X(t)\in \mathcal{R}^n$ be a non trivial solution to the following system of first order differential equations, called flow,
\begin{equation}\label{ds11}
\frac{\mathrm{d}\vec X}{\mathrm{d}t}=g(\vec X)\, ,
\end{equation}
here $g(\vec X)$ is a locally Lipschitz, one-to one continuous map
$g:\mathcal R^n\rightarrow \mathcal R^n$. Let $\vec X_*$ denotes the location of the fixed
points of the dynamical system (\ref{ds11}), and the corresponding
Jacobian matrix, which we denote as $\mathcal{J}(g)$, is equal to,
\begin{equation}\label{jaconiab}
(\mathcal{J})_{ij}=\Big{[}\frac{\mathrm{\partial g_i}}{\partial
X_j}\Big{]}\, .
\end{equation}
In order to have stable fixed points for system (\ref{ds11}) it is enough to set all eigenvalues of the Jacobian matrix so that $\lambda_i$ satisfies $\mathrm{Re}(\lambda_i)\neq 0$. The Hartman theorem predicts the
existence of a homeomorphism $\mathcal{F}:U\rightarrow \mathcal{R}^n$, where
$U$ is an open neighborhood of $\vec X_*$, such that
$\mathcal{F}(\vec X_*)$. The homeomorphism generates a flow
$\frac{\mathrm{d}h(u)}{\mathrm{d}t}$, which is,
\begin{equation}\label{fklow}
\frac{\mathrm{d}h(u)}{\mathrm{d}t}=\mathcal{J}h(u)\, ,
\end{equation}
It is proved that (\ref{fklow}) 
is a topologically conjugate flow to the one
system given in Eq. (\ref{ds11}). 

\subsection{Building the cosmological autonomous system of equations}
Now we study a model of interacting matter contents, where the continuity equation for each energy density $\rho_a$ is given by the following form:
\begin{eqnarray}
&&\dot{\rho}_a+3H(1+w_a)\rho_a=\Gamma_a\label{rhoeq}. 
\end{eqnarray}
where $a=\{m,\Lambda,r,\phi\}$ and $\Gamma_a$ is the interaction function given by the general form $\Gamma_a=\Gamma_a(\Omega_m,\Omega_{\Lambda},\Omega_r,\Omega_{\phi})$ and it satisfies $\Sigma_{a=1}^4\Gamma_a=0$. In $f(T)$ gravity, such interacting models are widely studied in the literatures, namely \cite{Jamil:2012nma}-\cite{Jamil:2012yz}. In Ref.\cite{arXiv:1012.4879}  authors showed that  the total gravitational energy is transferred from dark matter $\rho_m$ to dark energy $\rho_{\Lambda}$, and the cosmological coincidence problem in the Lambda-Cold Dark Matter ($\Lambda$CDM) model is slightly  assuaged.

In comparison to matter, DE and radiation energy densities, let us define an auxiliary scalar energy density as 
$$\Omega_{\phi}=\frac{A\dot{\phi}^2}{3H^2}$$
It is important to mention here that the auxiliary field $\phi$ is not a physical field. Consequently the kinetic term could be treated as tachyonic field as well as pressureless dust matter. In this paper we consider $w_{\phi}$ as a free parameter to be adjusted using observational data.

In this case we can write the following equation for the ratio between pressure and density , called effective  EoS equation, as follows:
\begin{widetext}
\begin{eqnarray}
&&\frac{2\dot{H}}{3H^2}=\frac{1}{2B-1}\Big(\Omega_{\phi}+
\Omega_m+(1+w_{\Lambda})\Omega_{\Lambda}+(1+w_{r})\Omega_r\Big)\label{hdot}.
\end{eqnarray}
\end{widetext}
where we prescribed the form of $f(\phi)$ as it is given in Eq. (\ref{fphi}) and 
we supposed that  $w_m=w_{\phi}=0,w_r=\frac{1}{3},w_{\Lambda}\in(-1,-\frac{1}{3})$. 

\par It is easy to rewrite  (\ref{rhoeq}) using the definition of 
\begin{eqnarray}
\Omega_a=\frac{\kappa \rho_a}{3H^2}\label{omegaa}
\end{eqnarray}
in the following set of first order differential equations where we used (\ref{hdot}) in it,
\begin{widetext}
\begin{eqnarray}\label{sys1}
&&\frac{d\Omega_a}{dN}=\frac{\kappa\Gamma_a}{3H^3}-\Omega_a\Big((1+w_a)+\frac{3}{2B-1}\Big(\Omega_{\phi}+
\Omega_m+(1+w_{\Lambda})\Omega_{\Lambda}+(1+w_{r})\Omega_r
\Big)\Big).
\end{eqnarray}
\end{widetext}
Recall $a=\{m,\Lambda,r,\phi\}$ and we use slow-roll variable $N=\log(\frac{a}{a_0})=-\ln(1+z)$ (the derivatives will be taken with respect to $N$) and  $z$ is redshift.
This is an autonomous system and should be analyzed in the vicinity of critical points where
$\frac{d\Omega_a}{dN}|_{c}=0$ using techniques developed in Sec. (\ref{Hartman-Grobman}).
\par
In terms of the variables (\ref{omegaa}) the Friedmann equation (\ref{eq11}) becomes the restriction :
\begin{eqnarray}
&& \Omega_m+\Omega_{\Lambda}+\Omega_r+\Omega_{\phi}=1-2B\label{frw1}
\end{eqnarray}

Note that due to the interaction term in the model, the density parameters $\Omega_m,\Omega_{\Lambda},\Omega_r,\Omega_{\phi}$  should be interpreted very strictly as effective density parameters. We mention here that the above constraint guaranteed the existence of possible cosmological attractors, because actually the shape of density functions remains typically the same and the full 4-dimensional configuration space constructed using density functions defines a shape invariant manifold and it defines the attractor solution in the dynamical system.

The effective EoS for system is defined 
\begin{widetext}
\begin{eqnarray}
	&&w_{eff}=\frac{p_{tot}}{\rho_{tot}}=-1-\frac{1}{2B-1}\Big(\Omega_{\phi}+
	\Omega_m+(1+w_{\Lambda})\Omega_{\Lambda}+(1+w_{r})\Omega_r\Big)\label{eos}.
\end{eqnarray} 
\end{widetext}
\section{Interaction term and phase portrait}
The general linear dependent model for interaction could be in the following form:
\begin{eqnarray}
&&\Gamma_a=\frac{3H^3}{\kappa}\Sigma_{b=1}^{4}\alpha_{ab}\Omega_b\label{Omega_b}.
\end{eqnarray}

There are some
criticisms about interacting models of DE, however the thermal properties of this
model in various gravities have been discussed in the literature
 \cite{interaction}. Furthermore, in Ref. \cite{He:2010im}, the authors proposed a systematic scheme to construct the
interaction form $\Gamma_a$  in a self consistent manner both in the perturbed form and in the background. They proved that in the perturbation formalism , there are
possible ways to break the degeneracy between the interaction, DE EoS and DM abundance.

With this choice, the system of equations (\ref{sys1})  are written in the following form:
\begin{widetext}
\begin{eqnarray}
&&\frac{d\Omega_a}{dN}=\Sigma_{b=1}^{4}\alpha_{ab}\Omega_b-\Omega_a\Big((1+w_a)+\frac{3}{2B-1}\Big(\Omega_{\phi}+
\Omega_m+(1+w_{\Lambda})\Omega_{\Lambda}+(1+w_{r})\Omega_r
\Big)
\Big)\label{model}
\end{eqnarray}
\end{widetext}
Let us study a class of these models where the DE interacts with both matter $\Omega_m$ and scalar field components. .
\par
Based on our former notation given in (\ref{Omega_b}), our interaction model is parametrized as follows:
\begin{eqnarray}
&&\alpha_{m\Lambda}=-6b,\ \  \alpha_{\Lambda\Lambda}=\alpha_{r\Lambda}=\alpha_{\phi\Lambda}=2b
\end{eqnarray}
The autonomous system of first order differential equations for density functions are written in the following forms,
\begin{widetext}
\begin{eqnarray}\label{sys111}
&&\frac{d\Omega_m}{dN}=-6b\Omega_{\Lambda}-\Omega_m\Big(1+\frac{3}{2B-1}\Big(\Omega_{\phi}+
\Omega_m+(1+w_{\Lambda})\Omega_{\Lambda}+(1+w_{r})\Omega_r
\Big)\Big)\equiv f_1\label{f1}\\&&
\frac{d\Omega_{\Lambda}}{dN}=2b\Omega_{\Lambda}-\Omega_{\Lambda}\Big((1+w_{\Lambda})+\frac{3}{2B-1}\Big(\Omega_{\phi}+
\Omega_m+(1+w_{\Lambda})\Omega_{\Lambda}+(1+w_{r})\Omega_r
\Big)\Big)\equiv f_2\label{f2}
\\&&
\frac{d\Omega_{r}}{dN}=2b\Omega_{\Lambda}-\Omega_{r}\Big(\frac{4}{3}+\frac{3}{2B-1}\Big(\Omega_{\phi}+
\Omega_m+(1+w_{\Lambda})\Omega_{\Lambda}+(1+w_{r})\Omega_r
\Big)\Big)\equiv f_3\label{f3}
\\&&
\frac{d\Omega_{\phi}}{dN}=2b\Omega_{\Lambda}-\Omega_{\phi}\Big((1+w_{\phi})+\frac{3}{2B-1}\Big(\Omega_{\phi}+
\Omega_m+(1+w_{\Lambda})\Omega_{\Lambda}+(1+w_{r})\Omega_r
\Big)\Big)\equiv f_4\label{f4}
\end{eqnarray}
\end{widetext}
These equations are related to the dynamics  and the interaction form, characterizing the main properties of  our model.
\subsection{The critical(fixed) points}

We stress here  that the high dimensionality of the phase space, where the system is described using dynamical systems presented in previous section, restricts us to have an effective graphical description of the phase space, and thus we will focus our investigations only on the analytical results.
\par
To make the dynamical analysis we first need to find the critical(fixed) points of
the system by setting the left hand side of equations (\ref{f1})-(\ref{f4}) to zero. Then we use the Hartman theorem to find the type and stability of each point \cite{dynamics}.


The location of the fixed
points $ P=
(\Omega_m,\Omega_{\Lambda},\Omega_r,\Omega_{\phi})
$ 
and their corresponding eigenvalues
of the dynamical system are in the following table,
where the stability of the fixed points is determined by evaluating the eigenvalues of the Jacobian matrix associated with the system.

\begin{widetext}
\begin{center}
\begin{tabular}{| c || c | c | c | c | c | c | c | c |}
\hline
P & $\Omega_m^c$ & $\Omega_{\Lambda}^c$ & $\Omega_r^c$ & $\Omega_{\phi}^c$ & $\lambda_1$ & $\lambda_2$ & $\lambda_3$ & $\lambda_4$\\ \hline \hline
 $A$ & $-\frac{2}{3} w_\phi B-\frac{1}{3}+\frac{1}{3} w_\phi$ & $0$ & $0$ & $0$ & $w_{\phi}$ & $w_{\phi} +1$ & $w_{\phi} - \frac{1}{3}$ & $2b-w_{\Lambda}+w_{\phi}$ \\ \hline
 $B$ & $0$ & $-\frac{2}{3} B+\frac{1}{3}$ & $0$ & $0$ & $-\frac{1}{3}$ & $1$ & $-w_{\phi}$ & $2b-w_{\Lambda}$ \\ \hline
 $C$ & $0$ & $0$ & $0$ & $0$ & $-\frac{4}{3}$ & $-1$ & $-w_{\phi} - 1$ & $2b-w_{\Lambda}-1$ \\ \hline
 $D$ & $0$ & $0$ & $0$ & $-\frac{2}{3} B+\frac{1}{3}$ & $\frac{1}{3}$ & $\frac{4}{3}$ & $-w_{\phi} + \frac{1}{3}$ & $\frac{1}{3}+2b-w_{\Lambda}$ \\ \hline
 $E$ & $x_1/x_2$ & $y_1/y_2$ & $z_1/x_2$ & $u_1/y_2$ & $\frac{1}{3}+2b-w_{\Lambda}$ & $-2b+w_{\Lambda}$ & $-2b+w_{\Lambda} +1$ & $-2b+w_{\Lambda}-w_{\phi}$ \\ \hline
\end{tabular}
\end{center}
\end{widetext}
where \\
$x_2=(-9w_{\Lambda}^4+(54b+9w_{\phi}-6)w_{\Lambda}^3 
+(-108b^2+(-36w_{\phi}+30)b+6w_{\phi}+3)w_{\Lambda}^2+(72b^3+(36w_{\phi}-48)b^2-3w_{\phi})w_{\Lambda} +24b^3+(-24w_{\phi}-12)b^2-12bw_{\phi})$ \\
$y_2=(-3w_{\Lambda}^4+(18b+3w_{\phi}-2)w_{\Lambda}^3+(-36b^2+(-12w_{\phi}+10)b+2w_{\phi}  +1)w_{\Lambda}^2+(24b^3+(12w_{\phi}-16)b^2-w_{\phi})w_{\Lambda}+8b^3+(-8w_{\phi}-4)b^2-4bw_{\phi})$ \\
$x_1=x(2b)(6b-3w_{\Lambda}+1)(2b-w_{\Lambda})$ \\  
$y_1=x(-2b)(6b-3w_{\Lambda}+1)(2b-w_{\Lambda} +w_{\phi})$ \\
$z_1=x(2b-w_{\Lambda})(2b-w_{\Lambda}+w_{\phi})(6b-3w_{\Lambda}+1)$\\ 
$u_1=x(2b)(2b-w_{\Lambda}))(2b-w_{\Lambda}+w_{\phi})$  \\
and \\
$x=(2b-w_{\Lambda}-1)(2B-1)$ \\
The corresponding
Jacobian matrix, which we denote as $\mathcal{J}(g)$, is equal to, \\
\[
\begin{bmatrix}
    \frac{\partial f_1}{\partial \Omega_m}        & \frac{\partial f_1}{\partial \Omega_{\Lambda}}& \frac{\partial f_1}{\partial \Omega_r} & \frac{\partial f_1}{\partial \Omega_{\phi}} \\
    \frac{\partial f_2}{\partial \Omega_m}        & \frac{\partial f_2}{\partial \Omega_{\Lambda}}& \frac{\partial f_2}{\partial \Omega_r} & \frac{\partial f_2}{\partial \Omega_{\phi}} \\
    \frac{\partial f_3}{\partial \Omega_m}        & \frac{\partial f_3}{\partial \Omega_{\Lambda}}& \frac{\partial f_3}{\partial \Omega_r} & \frac{\partial f_3}{\partial\Omega_{\phi}} \\
   \frac{\partial f_4}{\partial\Omega_m}        & \frac{\partial f_4}{\partial\Omega_{\Lambda}}& \frac{\partial f_4}{\partial\Omega_r} & \frac{\partial f_4}{\partial \Omega_{\phi}}
\end{bmatrix}
\]


The Eigenvalues and their stability for each point are written as following:

\begin{itemize}
\item Stability for point $A$: \\
The enough and sufficient condition to have $A_1$ as a stable fixed point for system  is that all eigenvalues of Jacobian matrix $\lambda_i$ must satisfy $\mathrm{Re}(\lambda_i)\neq 0$, i.e.,
 \begin{eqnarray}
 &&
 {w_{\Lambda} \le 2b-1,\ \  b < \frac{1}{3}w_{\Lambda}-\frac{1}{3}w_{\phi}}\\&&
 {w_{\phi} < -1,\ \  2b+\frac{1}{3} < w_{\Lambda}}\\&& {w_{\Lambda} \le 2b+\frac{1}{3},\ \  w_{\phi} < -1,\ \  2b-1 < w_{\Lambda}}
 \end{eqnarray}
 From these it is found that the stability occurs at: \\
 $2b-1 < w_{\Lambda} \le 2b+\frac{1}{3}$ \\
 $w_\phi < -1$. The corresponding effective EoS behaves like $w_{eff}=w_{\phi}$. Depending on the $w_{\phi}$, EoS evolves from larger than $-1$ to less than 
 $-1$, that is, it crosses the phantom divide line of $w_{eff} = -1$.
 

\item Stability for point $B$ \\
This is unstable critical point  and the corresponding effective EoS, $w_{eff}=w_{\Lambda}$ is always larger than $-1$ and  it crosses the phantom divide line of when $w_{\Lambda}=-1$.

\item Stability for point $C$: \\
Stability condition is  $-1 < w_{\phi}, 2b-1 < w_{\phi}$. Consequently $C$ can be stable conditionally. The corresponding effective EoS is  $w_{eff}=-1$ is located at the crosses the phantom divide line. 

\item Stability for point $D$: \\
We obviously conclude that is unstable. The corresponding effective EoS,is given $w_{eff}=0$ and is always larger than $-1$ and  it can not crosses the phantom divide line. 

\item Stability for point $E$: \\
The point is stable conditionally only and only if  $w_{\Lambda} < 2b-1, \frac{1}{2}w_{\Lambda}-\frac{1}{2}w_{\phi} < b$ . 

\end{itemize}

\par
\section{Cosmography}
The following types of observational data are commonly used to study cosmography,     
\noindent\textbf{ SNe Ia}: Type Ia supernovae (SNe Ia) or  the latest ``joint light curves" (JLA) sample \cite{sn}, comprised of 740 type Ia supernovae 
in the redshift range  $0.01 \leq z \leq 1.30$.
\\
\noindent\textbf{BAO}: The baryon acoustic oscillations (BAO)  \cite{bao1},\cite{bao2}, \cite{bao3}, and  \cite{bao4} (see table I of \cite{baotot}).
\\
\noindent\textbf{CC+$H_0$}:  The cosmic  chronometers (CC) data set  in the redshift range $0 < z < 2$ \cite{cc}. 
In $f(T)$ gravity cosmography introduced and invetigated in details in Ref. \cite{Capozziello:2011hj}.


 In Fig. 1, we plot the time evolution of the density functions 
for $b=0.5,0.7,0.9$, $w_{\Lambda}= -1/3$, $w_{\phi}= 0$, 
where the horizontal axis shows $\log(1+z)$ and the vertical axis does the 
value of the density functions. 
We can observe that the density functions of matter $\Omega_m$ and the cosmological constant $\Omega_\Lambda$ increase in time, whereas the density functions of radiation $\Omega_r$ and $\Omega_\phi$ decrease in time. For low redshift values, $0 < z < 0.1$, the densities $\Omega_\phi$, $\Omega_r$ are monotonically increasing functions, but matter and cosmological constant density functions decrease. At the present redshift $z\sim 0$, $\Omega_\phi \sim \Omega_r$ are negligible in comparison to the matter and cosmological constant densities. This confirms our remarkable observation about the scalar field $\phi$ that it cannot play the role of DE. So, the density of the scalar field is almost negligible at the present time. 
At distinct values of the redshift shown as $z^{*}$, densities of matte, radiation and scalar field become equal, i.e. $\Omega_m \sim \Omega_r \sim \Omega_{\phi}$. This occurs at $z^{*}\approx 0.1$. Furthermore, there is an era when $z^{\dagger} \sim 0.4$ in which $\Omega_m \sim \Omega_\lambda$, shows another equilibrium among matter and cosmological constant. 
These behaviors of the density functions are compatible with the observations. 

\begin{figure}[htbp]
\includegraphics[width=8.0cm]{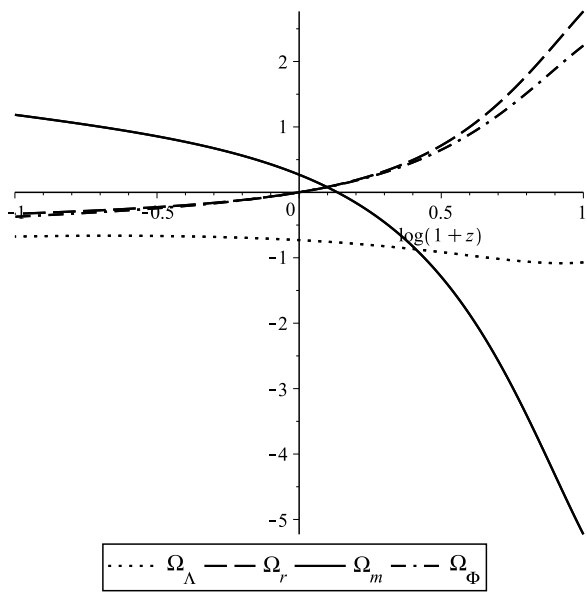}
%
\vspace{15mm}
  \caption{Numerical solutions for density functions for a set of parameters $b=0.5$ , $w_{\Lambda}= -1/3; w_{\phi}= 0$. Here, the horizontal axis shows $\log(1+z)$ and the vertical axis shows the value of the density functions.}
\end{figure}

\subsection{Observation of a type of deceleration to acceleration phase transition}
The deceleration parameter $q$ is defined as 
$$q=-1-\frac{\dot{H}}{H^2}.$$ If the expansion of the Universe is decelerating, $q >-1$, while if the cosmic expansion is accelerating, $q < -1$. In our model using (\ref{hdot}) we obtain 

\begin{eqnarray}
&&
q=-1-\frac{3}{2(2B-1)}\Big(\Omega_{\phi}+
\Omega_m\\&&\nonumber+(1+w_{\Lambda})\Omega_{\Lambda}+(1+w_{r})\Omega_r\Big)\label{q}.
\end{eqnarray}

We find numerical solutions by using $h = 0.7127_{-0.015}^{+0.013}$ km/s/Mpc, $\Omega_{\Lambda }=0.7018_{-0.02}^{+0.018}$, and  $\Omega_{m0}=0.2981_{-0.018}^{+0.02}$, 
with $\chi^2_{min}=707.4$, $H_0= 73. 24 \pm 1.74$ km/s/Mpc.
In Fig. 2, we depict the time evolution of the deceleration parameter $q$ 
for $b=0.5, 0.7, 0.9$. Here, the horizontal axis shows $\log(1+z)$ and the vertical axis shows the value of $q$. 
All of the curves meet. From Fig. 2, it is found that in the past for lower values of redshift, 
the value of $q$ evolved from larger than $-1$ to less than $-1$, 
namely, the expansion phase of the Universe changed from 
the deceleration to the acceleration. This is consistent with the observations. 

\begin{figure}
     	\includegraphics[width=8.0cm]{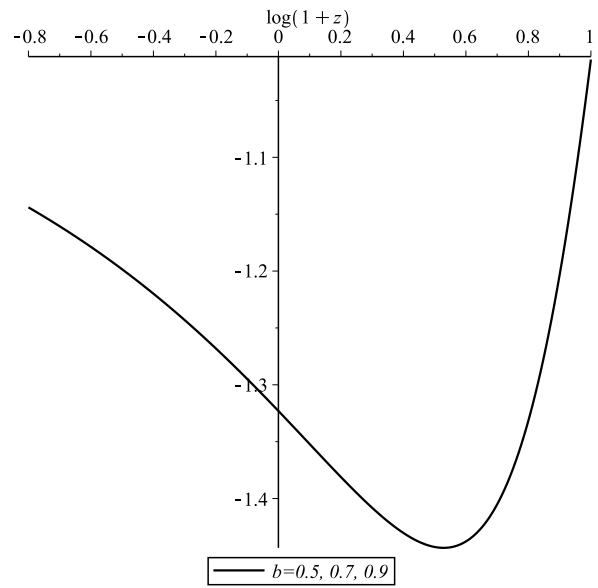}
\vspace{15mm}
     	\caption{\label{q} Deceleration parameter $q$ for $b=0.5, 0.7, 0.9$. All of the curves meet. Here, the horizontal axis shows $\log(1+z)$ and the vertical axis shows the value of $q$.}
\end{figure} 

\subsection{ Effective Equation of State (EoS) of the Universe}
Effective EoS of the Universe was defined in Eq. (\ref{eos}). 
We note that in the DE dominated stage, the value of 
the EoS of DE can be regarded as 
the effective EoS of the Universe $w_{eff}$. 

In Fig. 3, we show the time evolution of the effective EoS $w_{eff}$ 
for $b=0.5,0.7,0.9$ , 
where the horizontal axis shows $\log(1+z)$ and the vertical axis shows the value of 
$w_{eff}$. From Fig. 3, it is found that  for low redshift values , the value of 
$w_{eff}$ became less than $-\frac{1}{3}$ and therefore the cosmic 
expansion phase of the Universe changed from the deceleration 
to the acceleration. It is also seen that in our model, 
the value of $w_{eff}$ evolves from larger than $-1$ to less than $-1$; that is, it crosses the phantom divide line of $w_{eff} = -1$. The value of the Eos parameter at the present redshift is around $w_{eff}\sim -1.6<-1$ shows an acceleration expansion beyond the phantom line. 

\begin{figure}
\includegraphics[width=8.0cm]{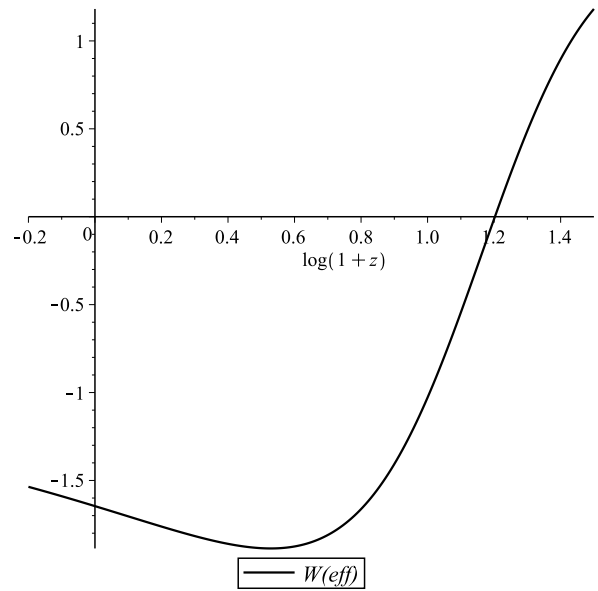}
\vspace{15mm}
     	\caption{\label{weffl} Effective EoS $w_{eff}$ for  $b=0.5,b=0.7,b=0.9$. The three curves coincides. Here, the horizontal axis shows $\log(1+z)$ and the vertical axis shows the value of $w_{eff}$.}
\end{figure}

\section{de Sitter solution}
Cosmological models usually have de Sitter (dS) solution where the Hubble parameter is constant (or almost constant in inflationary scenarios) $H=H_0$ as trivial solution. In GR such solution (used in inflationary mechanism as well as late time cosmology) becomes accessible when the dominant energy density  $\rho\approx\rho_0$, which means that to have dS  we need matter fields with very slowly varying energy density. It is not possible to find dS solution as an empty space  solution in GR . But in modified gravity because of   the geometrical terms (curvature $\mathcal{R}$ or torsion $T$) it will be possible to find dS as an (almost) exact solution for field equations.  In forthcoming sections, we look for dS solution both in empty and matter contents cases in model defined by Eqs. (\ref{eq1},\ref{eq1}).\par
 Let us firstly perform a little investigation on the equations of motion. As a result of continuity eq, we have two additional Klein-Gordon like dissipative eqs. (\ref{rhoeq}) in non interacting case, we can find equations of motion for $\phi,\xi$. If we suppose that in (\ref{fphi}), $B=0,A=-1$, they are written as following:
\begin{eqnarray}
&&\ddot{\phi}+3H\dot{\phi}+6H^2=0\label{phi},\\
&&\ddot{\xi}+3H\dot{\xi}+6H^2=0\label{xi}.
\end{eqnarray}
Note that always with $f(\phi)=-\phi$, we have $\nabla^{\mu}\nabla_{\mu}(\phi-\xi)=0$, and we have a "gauge freedom" to write fields $\phi,\xi$ as follows:

\begin{eqnarray}
&&\phi-\xi=\eta(t)\label{gauge}.
\end{eqnarray}
An exact solution for $\eta$, in FLRW background is given by,
\begin{eqnarray}
&&\eta(t)=\int\frac{\eta_0}{a(t)^3} dt\label{soleta}.
\end{eqnarray}
Note that $\dot{\eta}(t)\sim \rho_m(t)$. In this case we can interpret $\dot{\eta}(t)$ as cold dark matter density. It is possible to take $\eta_0=0$ or $\eta_0\neq0$. We will study both cases in next subsections.

\subsection{Empty spacetime}\label{vacuum}
To find an exact (almost exact) dS solution for a system given by (\ref{eq1}-\ref{eq2}),(\ref{phi}-\ref{xi}) in vacuum let us relax all matter contents, to make spacetime empty (we will  never consider a quantum fluctuations in this approach).  Furthermore, we set $H=H_0$ for dS case.
\subsubsection{Case $\eta(t)=0$}:
When $\eta=0$, $\phi=\xi$. Thus the system (\ref{phi},\ref{xi}) and (\ref{hdot}) is reduced to the following system:
\begin{eqnarray}
&&\frac{H_0}{\dot{\phi}}=\frac{1}{4}\\&& \ddot{\phi}+3H_0\dot{\phi}+6H_0^2=0.
\end{eqnarray}
From first equation we find $\phi(t)=4H_0t+\phi_0$. If we substitute it in second equation we obtain $H_0=0$. In $f(T)$ gravity stability for Einstein Universe is well studied in \cite{Wu:2011xa}.

This is just Einstein static Universe and it proves that no dS solution exist. \par
Let us check whether this solution is stable or not. We make perturbation around the solution given by $(H,\phi)=(0,\phi_0)$. The  equation is given by the following:
\begin{eqnarray}
&&\delta\ddot\phi+3\dot{\phi}\delta H +3H\delta\dot\phi +12H\delta H=0.
\end{eqnarray}
substituting the zeroth order solution we find
\begin{eqnarray}
&&\delta\ddot\phi=0.
\end{eqnarray}
Exact solution for perturbation function is 
\begin{eqnarray}
&&\delta\phi=at+b.
\end{eqnarray}
When $t\to\infty$, we clearly observe that perturbation is growing up linearly and consequently the system becomes unstable under infinitesimal field and background perturbations.

\subsubsection{Case $\eta(t)\neq0$}:
When $\eta= \int\frac{\eta_0}{a(t)^3} dt$, we have  $\dot{\phi}-\dot{\xi}=\frac{\eta_0}{a(t)^3} $. Thus the system (\ref{phi},\ref{xi}) and (\ref{hdot}) is reduced to the following system:
\begin{eqnarray}
&&\frac{1}{2H_0}=\frac{1}{\dot{\phi}}+\frac{1}{\dot{\xi}},\\&&
\dot{\phi}-\dot{\xi}=\frac{\eta_0}{a(t)^3}
\end{eqnarray}
Note that in dS phase, $a(t)=a_0 e^{H_0 t}$. An exact solution for fields pair $\phi,\xi$ is given as follows:

\begin{eqnarray}
&&\phi_0(t)=C_2+2H_0t-\frac{\eta_0}{6a_0^3H_0}e^{-3H_0 t}\\&&\nonumber
+\frac{1}{6H_0a_0^3}\Big(\Delta-4a_0^3\sqrt{H_0}\tanh^{-1}(\frac{\Delta}{4a_0^3\sqrt{H_0}})
\Big)
\label{solphi1}\\
&&\xi _0( t) =C_2+C_1+2H_0t+\frac{\eta_0}{6a_0^3H_0}e^{-3H_0 t}\\&&\nonumber
+\frac{1}{6H_0a_0^3}\Big(\Delta-4a_0^3\sqrt{H_0}\tanh^{-1}(\frac{\Delta}{4a_0^3\sqrt{H_0}})\Big)
\label{solxi1}
\end{eqnarray}

Here $C_1,C_2$ are arbitrary integration constants and $\Delta=16H_0 a_0^6+a_0^3H_0^2e^{-6H_0t}$.
The solutions given in (\ref{solphi1},\ref{solxi1}) are exact solutions for dS phase of our model under study. Let us study its stability under perturbations in time representation of fields and backgrounds. Later in Sec. (\ref{Stability in vacuum}) another equivalent  analysis using slow-roll coordinate N will be introduced. \par
Perturbation of field equations (\ref{phi},\ref{xi}) and Eq. (\ref{eq2}) around the solution given by $(H,\phi,\xi)=(H_0,\phi_0,\xi_0)$ substituted in Eqs.(\ref{solphi1},\ref{solxi1}) given the following system of Eqs:

\begin{eqnarray}
&&\delta\ddot\phi+3\dot{\phi}_0\delta H +3H_0\delta\dot\phi +12H_0\delta H=0\\
&&\delta\ddot\xi+3\dot{\xi}_0\delta H +3H_0\delta\dot\xi +12H_0\delta H=0\\&&
2\delta\dot H +4\phi_0\delta\dot{H}+(\dot{\xi}_0-2H_0)\delta\dot{\phi}+
\\&&\nonumber(\dot{\phi}_0-2H_0)\delta\dot{\xi}-2(\dot{\xi}_0+\dot{\phi}_0)\delta H=0.
\end{eqnarray}

It is hard to find exact solutions for perturbation functions and we do not discuss it. Note that asymptotically, $\phi\sim \xi\approx 2H_0 t$, consequently we can find the following solutions which are valid only when $t\to \infty$:

\begin{eqnarray}
&&\delta\phi\sim \delta\xi\approx (H_0t)^2,\ \ \delta H\approx H_0t.
\end{eqnarray}
We conclude that dS behaves as an unstable phase in our model.

\subsection{Case of matter contents}\label{nonvacuum}
Now we study exact solutions in dS phase when $\rho_a\neq0$. As a general case we consider the model given in (\ref{eq1},\ref{eq2}) for general densities $\rho_m=\rho_m^{0}a(t)^{-3}$. 
\subsubsection{Case $\eta(t)=0$}
When $\eta=0$, $\phi=\xi$, the system (\ref{phi},\ref{xi}) and (\ref{hdot}) is reduced to the following system:
\begin{eqnarray}
&&\dot{\phi}^2-4H_0\dot{\phi}+\kappa\rho_m^{0}a(t)^{-3}
=0.\label{hdot2}\\
&&\ddot{\phi}+3H_0\dot{\phi}+6H_0^2=0.
\end{eqnarray}
Exact solutions provide that $H_0=0$ is the only possible solution. Thus similar to the empty case, still we just have Einstein static Universe. \par
If We perturb the system around the solutions given above, we have the following system of equations:

\begin{eqnarray}
&&\delta\ddot\phi=0\\
&&
2(1+2\phi_0)\delta\dot H -\delta \rho_m(t)=0.
\end{eqnarray}
where $\phi_0=\mbox{C}$ is a constant. The system is clearly asymptotically unstable, consequently no stable static Einstein Universe exists.

\subsubsection{Case $\eta(t)\neq0$}
When $\eta= \int\frac{\eta_0}{a(t)^3} dt$,  the system (\ref{phi},\ref{xi}) and (\ref{hdot}) is reduced to the following system:
\begin{eqnarray}
&&\dot{\phi}\dot{\xi}-2H_0(\dot{\phi}+\dot{\xi})+\kappa\rho_m^{0}a(t)^{-3}
=0.\label{hdot3},\\&&
\dot{\phi}-\dot{\xi}=\frac{\eta_0}{a(t)^3}
\end{eqnarray}
Exact solutions are given in terms of first integrals:
\begin{eqnarray}
&&\dot{\phi}=2H_0+\frac{\eta_0}{2a(t)^3}\\&&\nonumber-\frac{1}{2}\sqrt{\eta_0a(t)^{-6}-4\kappa\rho_m^{0}a(t)^{-3}+16H_0^2
}\\&&
\dot{\xi}=2H_0-\frac{\eta_0}{2a(t)^3}\\&&\nonumber-\frac{1}{2}\sqrt{\eta_0a(t)^{-6}-4\kappa\rho_m^{0}a(t)^{-3}+16H_0^2
}.
\end{eqnarray}
Therefore we have dS solution. We also conclude here that the system behaves asymptotically, $\phi\sim \xi\approx 2H_0 t$, consequently we can find the perurbated  solutions when $t\to \infty$ diverge and consequently system becomes unstable.

\section{Stability of de Sitter via Hartman-Grobman linearizion theorem }
Stability of dS solution plays an essential role in inflationary scenario to have thermalization phase. Because nonlocal theory supposed to be an alternative theories for inflation we will study the stability in this context.

\subsection{Stability in vacuum}\label{Stability in vacuum}
In vacuum we observed that only when $\eta\neq0$ we have dS solutions given by  (\ref{solphi1},\ref{solxi1}). Let us see whether this solution is stable or not. Suppose that $H=H_0,$ and  (\ref{solphi1},\ref{solxi1}) are exact solutions for (\ref{eq1},\ref{eq2}) and (\ref{phi},\ref{xi}) with $f(\phi)=-\phi$. We can find the following auxiliary system:
\begin{eqnarray}
&&\ddot{\xi}+3H\dot{\xi}+6H^2=0\label{ds1}\\&&
\ddot{\phi}+3H\dot{\phi}+6H^2=0\label{ds2}\\&&
\frac{\dot{H}}{3H^2}=2H(\frac{1}{\dot{\phi}}+\frac{1}{\dot{\xi}})-1\label{ds3}.
\end{eqnarray}
Note that Eq. $\dot{\phi}-\dot{\xi}=\frac{\eta_0}{a(t)^3}$ is obtained using subtraction of two KG eqs for $\phi,\xi$.
To study stability the first step is to make system given in (\ref{ds1}-\ref{ds3}) dimensionless, using new time parameter $N$ and new derivative $'=\frac{d}{dN}$ we have:
\begin{eqnarray}
&&\varphi'=-6(2+\frac{\phi}{\alpha})\label{varphi}\\&&
\alpha'=-6(2+\frac{\alpha}{\varphi})\label{alpha}\\&&
H'=-3H(1-\frac{2}{\varphi}-\frac{2}{\alpha})\label{Hprime}.
\end{eqnarray}
where $\varphi\equiv \phi',\alpha\equiv \xi'$. The critical point is located at $A=(H=H_0,\alpha=0,\varphi=0)$. We linearize the system under perturbation functions $H=H_0+\delta H, \varphi=\delta\varphi, \alpha=\delta\alpha$, The corresponding Matrix has an eigenvalue $\lambda_1=0$, and it shows that dS solution is an unstable point.
\subsection{Stability in matter mixture}\label{Stability in matter mixture}
FLRW equations are given as following forms for the case with matter density and when $\phi\neq \xi$:
\begin{eqnarray}
&&\frac{2\dot{H}}{3H^2}=\frac{-\dot{\phi}\dot{\xi}+2H(\dot{\phi}+\dot{\xi})-\kappa\rho_m^{0}a(t)^{-3}}{\frac{1}{2}\dot{\phi}\dot{\xi}+\kappa\rho_m^{0}a(t)^{-3}}\\&&
\ddot{\xi}+3H\dot{\xi}+6H^2=0\\&&
\ddot{\phi}+3H\dot{\phi}+6H^2=0.
\end{eqnarray}

In this case because of $a(t)$ the system becomes non autonomous but still we can study the local stability in the vicinity of a critical point for $t\geq t_s$. Using the time coordinate $N$ and by redefining  $\varphi\equiv \phi',\alpha\equiv \xi'$ we have the following system of differential equations:
\begin{widetext}
\begin{eqnarray}
&&\frac{2}{3}\frac{H'}{H}=\frac{H^2(2(\varphi+\alpha)-\varphi\alpha)-\kappa\rho_m^{0}a^{-3}}
{\frac{1}{2}H^2\varphi\alpha+\kappa\rho_m^{0}a^{-3}}\\&&
\varphi'=\,{\frac {-3(4\,{H}^{2}{a}^{3}\alpha\,\varphi+2\,{H}^{2}{a}^{3}{\varphi}^{2}+
\kappa\,\varphi\,\rho^{0}_{{m}}+4\,\kappa\,\rho^{0}_{{m}})}{{H}^{2}{a}^{3}\alpha\,
\varphi+2\,\kappa\,\rho^{0}_{{m}}}}
\\&&
\alpha'=\,{\frac {-3(2\,{H}^{2}{a}^{3}{\alpha}^{2}+4\,{H}^{2}{a}^{3}\alpha\,
\varphi+\alpha\,\kappa\,\rho^{0}_{{m}}+4\,\kappa\,\rho^{0}_{{m}})}{{H}^{2}{a}^{3}
\alpha\,\varphi+2\,\kappa\,\rho^{0}_{{m}}}}
\end{eqnarray}
\end{widetext}
here $\varphi\equiv \phi',\alpha\equiv \xi'$.

The corresponding linearized system near the unique physically accepted critical point $\{H = 0,\alpha = -4, \phi = -4\}$ called  a proper node, or a
star point(Actually it defines Einstein static solution) has a triplet proper node degenerated eigen value $\lambda=-\frac{3}{2}$. The critical point is asymptotically stable , and  it shows that system is  asymptotically stable .

\section{Conclusions}
In this paper, we have considered the exact cosmological solutions in nonlocal $f(T)$ gravity, which can be regarded as an extension of nonlocal general relativity to the Weitzenb\"{o}ck spacetime. 
We have explored the autonomous system of the first order dynamical equations 
by deriving effective forms of cosmological field equations in a homogeneous and isotropic cosmological background to describe the whole evolution history of the universe. 
Furthermore, we have introduced a specific form of the interaction between matter, DE, radiation and a scalar field and analyzed the local stability in the dynamical systems, which is the so-called ``the stable/unstable manifold''. It has been found that the system has a stable(unstable) attractor solutions. 
In addition, we have investigated the exact solutions of the cosmological equations in the case of de Sitter spacetime. We have demonstrated whether the de Sitter solutions can exist or not in this scenario by examining the role of an auxiliary function called ``gauge'' $\eta$ in the formation of such cosmological solutions. 
Moreover, we have studied the stability problem of the de Sitter solutions both in vacuum and non-vacuum spacetimes. It has been shown that for nonlocal $f(T)$ gravity, we can obtain the stable de Sitter solutions even in vacuum spacetime.

\begin{acknowledgments}
We thank the anonymous referee for intuitive comments
and thorough criticism on our manuscript. This
work was partially supported by the JSPS KAKENHI
Grant Number JP 25800136 and Competitive Research Funds for Fukushima University Faculty (17RI017 and 18RI009) (K.B.).
\end{acknowledgments}


\begin{thebibliography}{99}


\bibitem{Inflation} 
%
  A.~A.~Starobinsky,
  Phys.\ Lett.\ B {\bf 91}, 99 (1980);
%
  K.~Sato,
  Mon.\ Not.\ Roy.\ Astron.\ Soc.\  {\bf 195}, 467 (1981); 
%
  A.~H.~Guth,
  Phys.\ Rev.\ D {\bf 23}, 347 (1981);
%
  A.~D.~Linde,
  Phys.\ Lett.\ B {\bf 108}, 389 (1982);
%
  A.~Albrecht and P.~J.~Steinhardt,
  Phys.\ Rev.\ Lett.\  {\bf 48}, 1220 (1982). 
%

\bibitem{SN}
%
  S.~Perlmutter {\it et al.} [Supernova Cosmology Project Collaboration],
  Astrophys.\ J.\  {\bf 517}, 565 (1999)
  [astro-ph/9812133]; 
%
  A.~G.~Riess {\it et al.} [Supernova Search Team Collaboration],
  Astron.\ J.\  {\bf 116}, 1009 (1998)
  [astro-ph/9805201].

\bibitem{Ade:2015xua} 
  P.~A.~R.~Ade {\it et al.} [Planck Collaboration],
  Astron.\ Astrophys.\  {\bf 594}, A13 (2016) 
  [arXiv:1502.01589 [astro-ph.CO]];
  P.~A.~R.~Ade {\it et al.} [Planck Collaboration],
  Astron.\ Astrophys.\  {\bf 594}, A20 (2016) 
  [arXiv:1502.02114 [astro-ph.CO]];
  P.~A.~R.~Ade {\it et al.}  [BICEP2 Collaboration],
  Phys.\ Rev.\ Lett.\  {\bf 112}, 241101 (2014) 
  [arXiv:1403.3985 [astro-ph.CO]];
  P.~A.~R.~Ade {\it et al.}  [BICEP2 and Planck Collaborations],
  Phys.\ Rev.\ Lett.\  {\bf 114}, 101301 (2015)
  [arXiv:1502.00612 [astro-ph.CO]];
  P.~A.~R.~Ade {\it et al.} [BICEP2 and Keck Array Collaborations],
  Phys.\ Rev.\ Lett.\  {\bf 116}, 031302 (2016)
  [arXiv:1510.09217 [astro-ph.CO]];
E.~Komatsu {\it et al.} [WMAP Collaboration],
Astrophys.\ J.\ Suppl.\ {\bf 192}, 18 (2011) 
[arXiv:1001.4538 [astro-ph.CO]];
  G.~Hinshaw {\it et al.}  [WMAP Collaboration],
  Astrophys.\ J.\ Suppl.\  {\bf 208}, 19 (2013) 
  [arXiv:1212.5226 [astro-ph.CO]].

\bibitem{LSS}
%
  M.~Tegmark {\it et al.} [SDSS Collaboration],
  Phys.\ Rev.\ D {\bf 69}, 103501 (2004)
  [astro-ph/0310723]; \\
%
  U.~Seljak {\it et al.} [SDSS Collaboration],
  Phys.\ Rev.\ D {\bf 71}, 103515 (2005)
  [astro-ph/0407372].

\bibitem{Eisenstein:2005su} 
  D.~J.~Eisenstein {\it et al.} [SDSS Collaboration],
  Astrophys.\ J.\  {\bf 633}, 560 (2005)
  [astro-ph/0501171].

\bibitem{Jain:2003tba} 
  B.~Jain and A.~Taylor,
  Phys.\ Rev.\ Lett.\  {\bf 91}, 141302 (2003)
  [astro-ph/0306046].

\bibitem{R-DE-MG}
%
S.~Nojiri and S.~D.~Odintsov,
Phys.\ Rept.\ {\bf 505}, 59 (2011)
[arXiv:1011.0544 [gr-qc]];\\ 
%
S.~Nojiri and S.~D.~Odintsov,
eConf C {\bf 0602061} (2006) 06  
[Int.\ J.\ Geom.\ Meth.\ Mod.\ Phys.\ {\bf 4}, 115 (2007)]
[hep-th/0601213]; 
%
%
%
%
  K.~Bamba, S.~Capozziello, S.~Nojiri and S.~D.~Odintsov,
  Astrophys.\ Space Sci.\  {\bf 342}, 155 (2012)
  [arXiv:1205.3421 [gr-qc]];  
%
%
%
  K.~Bamba and S.~D.~Odintsov,
  Symmetry {\bf 7}, 1, 220 (2015)
  [arXiv:1503.00442 [hep-th]].


\bibitem{T-G}
%
  F.~W.~Hehl, P.~Von Der Heyde, G.~D.~Kerlick and J.~M.~Nester,
  Rev.\ Mod.\ Phys.\  {\bf 48}, 393 (1976);\\ 
  K.~Hayashi and T.~Shirafuji,
  Phys.\ Rev.\  D {\bf 19}, 3524 (1979)
  [Addendum-ibid.\  D {\bf 24}, 3312 (1982)];\\ 
  E.~E.~Flanagan and E.~Rosenthal,
  Phys.\ Rev.\  D {\bf 75}, 124016 (2007) 
  [arXiv:0704.1447 [gr-qc]];\\
%
  J.~Garecki,
  arXiv:1010.2654 [gr-qc].
%

\bibitem{Cai:2015emx} 
  Y.~F.~Cai, S.~Capozziello, M.~De Laurentis and E.~N.~Saridakis,
  Rept.\ Prog.\ Phys.\  {\bf 79}, 106901 (2016) 
  [arXiv:1511.07586 [gr-qc]].

\bibitem{F-R}
%
  H.~A.~Buchdahl,
  Mon.\ Not.\ Roy.\ Astron.\ Soc.\  {\bf 150}, 1 (1970);\\ 
%
  S.~Capozziello, S.~Carloni and A.~Troisi,
  Recent Res.\ Dev.\ Astron.\ Astrophys.\  {\bf 1}, 625 (2003)
  [astro-ph/0303041]; 
%
  S.~Nojiri and S.~D.~Odintsov,
  Phys.\ Rev.\ D {\bf 68}, 123512 (2003)
  [hep-th/0307288];\\ 
%
  S.~M.~Carroll, V.~Duvvuri, M.~Trodden and M.~S.~Turner,
  Phys.\ Rev.\ D {\bf 70}, 043528 (2004)
  [astro-ph/0306438].



\bibitem{F-T-Inf}
%
  R.~Ferraro and F.~Fiorini,
  Phys.\ Rev.\  D {\bf 75}, 084031 (2007)
  [arXiv:gr-qc/0610067];\\ 
%
  R.~Ferraro and F.~Fiorini,
  Phys.\ Rev.\  D {\bf 78}, 124019 (2008) 
  [arXiv:0812.1981 [gr-qc]];\\ 
%
  K.~Bamba, S.~Nojiri and S.~D.~Odintsov,
  Phys.\ Lett.\ B {\bf 731}, 257 (2014)
  [arXiv:1401.7378 [gr-qc]].
%

\bibitem{F-T-LC}
%
  G.~R.~Bengochea and R.~Ferraro,
  Phys.\ Rev.\  D {\bf 79}, 124019 (2009)
  [arXiv:0812.1205 [astro-ph]]; 
%
  E.~V.~Linder,
  Phys.\ Rev.\  D {\bf 81}, 127301 (2010)
  [Erratum-ibid.\  D {\bf 82}, 109902 (2010)]
  [arXiv:1005.3039 [astro-ph.CO]];
%
  K.~Bamba, C.~Q.~Geng, C.~C.~Lee and L.~W.~Luo,
  JCAP {\bf 1101}, 021 (2011)
  [arXiv:1011.0508 [astro-ph.CO]]; 
%
  K.~Bamba, C.~Q.~Geng and C.~C.~Lee,
  arXiv:1008.4036 [astro-ph.CO].
%

\bibitem{F(T)-Refs}
%
  P.~Wu and H.~W.~Yu,
  Phys.\ Lett.\  B {\bf 693}, 415 (2010) 
  [arXiv:1006.0674 [gr-qc]];\\ 
%
%
%
  P.~Wu and H.~Yu,
  Phys.\ Lett.\  B {\bf 692}, 176 (2010) 
  [arXiv:1007.2348 [astro-ph.CO]];\\ 
%
%
%
  S.~H.~Chen, J.~B.~Dent, S.~Dutta and E.~N.~Saridakis,
  Phys.\ Rev.\  D {\bf 83}, 023508 (2011) 
  [arXiv:1008.1250 [astro-ph.CO]];
%
  G.~R.~Bengochea,
  Phys.\ Lett.\  B {\bf 695}, 405 (2011) 
  [arXiv:1008.3188 [astro-ph.CO]];
%
  C.~Q.~Geng, C.~C.~Lee, E.~N.~Saridakis and Y.~P.~Wu,
  Phys.\ Lett.\ B {\bf 704}, 384 (2011)
  [arXiv:1109.1092 [hep-th]]; 
%
  K.~Bamba and C.~Q.~Geng,
  JCAP {\bf 1111}, 008 (2011)
  [arXiv:1109.1694 [gr-qc]];
  R.~Ferraro and F.~Fiorini,
  Phys.\ Rev.\ D {\bf 84}, 083518 (2011)
  [arXiv:1109.4209 [gr-qc]];
%
  H.~Wei,
  Phys.\ Lett.\ B {\bf 712}, 430 (2012)
  [arXiv:1109.6107 [gr-qc]];
%
  C.~Q.~Geng, C.~C.~Lee and E.~N.~Saridakis,
  JCAP {\bf 1201}, 002 (2012)
  [arXiv:1110.0913 [astro-ph.CO]];
%
  P.~A.~Gonzalez, E.~N.~Saridakis and Y.~Vasquez,
  JHEP {\bf 1207}, 053 (2012)
  [arXiv:1110.4024 [gr-qc]];
%
  K.~Bamba, R.~Myrzakulov, S.~Nojiri and S.~D.~Odintsov,
  Phys.\ Rev.\ D {\bf 85}, 104036 (2012)
  [arXiv:1202.4057 [gr-qc]];
%
  M.~Jamil, D.~Momeni and R.~Myrzakulov,
  Eur.\ Phys.\ J.\ C {\bf 72}, 1959 (2012)
  [arXiv:1202.4926 [physics.gen-ph]];
%
  M.~Jamil, D.~Momeni and R.~Myrzakulov,
  Eur.\ Phys.\ J.\ C {\bf 72}, 2075 (2012) 
  [arXiv:1208.0025 [gr-qc]];
%
  M.~Jamil, D.~Momeni and R.~Myrzakulov,
  Eur.\ Phys.\ J.\ C {\bf 72}, 2122 (2012) 
  [arXiv:1209.1298 [gr-qc]];
%
  M.~Jamil, D.~Momeni and R.~Myrzakulov,
  Eur.\ Phys.\ J.\ C {\bf 72}, 2137 (2012) 
  [arXiv:1210.0001 [physics.gen-ph]];
%
  K.~Izumi and Y.~C.~Ong,
  JCAP {\bf 1306}, 029 (2013)
  [arXiv:1212.5774 [gr-qc]];
%
  M.~Jamil, D.~Momeni and R.~Myrzakulov,
  Eur.\ Phys.\ J.\ C {\bf 73}, 2267 (2013) 
  [arXiv:1212.6017 [gr-qc]];\\ 
%
  K.~Bamba, S.~D.~Odintsov and E.~N.~Saridakis,
  Mod.\ Phys.\ Lett.\ A {\bf 32}, 1750114 (2017)
  [arXiv:1605.02461 [gr-qc]].
%

\bibitem{L-L-I}
%
  B.~Li, T.~P.~Sotiriou and J.~D.~Barrow,
  Phys.\ Rev.\  D {\bf 83}, 064035 (2011)
  [arXiv:1010.1041 [gr-qc]];\\ 
%
  T.~P.~Sotiriou, B.~Li and J.~D.~Barrow,
  Phys.\ Rev.\  D {\bf 83}, 104030 (2011)
  [arXiv:1012.4039 [gr-qc]]. 
%

\bibitem{RP-LLI}
%
  R.~Ferraro and F.~Fiorini,
  Phys.\ Lett.\  B {\bf 702}, 75 (2011)
  [arXiv:1103.0824 [gr-qc]];\\ 
%
  M.~Li, R.~X.~Miao and Y.~G.~Miao,
  JHEP {\bf 1107}, 108 (2011)
  [arXiv:1105.5934 [hep-th]];\\ 
%
  Y.~C.~Ong, K.~Izumi, J.~M.~Nester and P.~Chen,
  Phys.\ Rev.\ D {\bf 88}, 024019 (2013)
  [arXiv:1303.0993 [gr-qc]];
%
  K.~Bamba, S.~D.~Odintsov and D.~S\'{a}ez-G\'{o}mez,
  Phys.\ Rev.\ D {\bf 88}, 084042 (2013)
  [arXiv:1308.5789 [gr-qc]]; 
%
  K.~Bamba, S.~Capozziello, M.~De Laurentis, S.~Nojiri and D.~S\'{a}ez-G\'{o}mez,
  Phys.\ Lett.\ B {\bf 727}, 194 (2013)
  [arXiv:1309.2698 [gr-qc]];
%
  K.~Izumi, J.~A.~Gu and Y.~C.~Ong,
  Phys.\ Rev.\ D {\bf 89}, 084025 (2014)
  [arXiv:1309.6461 [gr-qc]]; 
%
  P.~Chen, K.~Izumi, J.~M.~Nester and Y.~C.~Ong,
  Phys.\ Rev.\ D {\bf 91}, 064003 (2015)
  [arXiv:1412.8383 [gr-qc]]; 
%
  S.~Bahamonde, C.~G.~B\"{o}hmer and M.~Wright,
  Phys.\ Rev.\ D {\bf 92}, 104042 (2015)
  [arXiv:1508.05120 [gr-qc]].
%

\bibitem{Deser:2007jk}
  S.~Deser and R.~P.~Woodard,
  Phys.\ Rev.\ Lett.\  {\bf 99}, 111301 (2007)
  [arXiv:0706.2151 [astro-ph]].

\bibitem{Nojiri:2007uq}
  S.~Nojiri and S.~D.~Odintsov,
  Phys.\ Lett.\  B {\bf 659}, 821 (2008)
  [arXiv:0708.0924 [hep-th]].

\bibitem{ArkaniHamed:2002fu}
  N.~Arkani-Hamed, S.~Dimopoulos, G.~Dvali and G.~Gabadadze,
  arXiv:hep-th/0209227.

\bibitem{Nojiri:2010pw}
  S.~Nojiri, S.~D.~Odintsov, M.~Sasaki and Y.~l.~Zhang,
  Phys.\ Lett.\  B {\bf 696}, 278 (2011)
  [arXiv:1010.5375 [gr-qc]].

\bibitem{Bamba:2012ky} 
  K.~Bamba, S.~Nojiri, S.~D.~Odintsov and M.~Sasaki,
  Gen.\ Rel.\ Grav.\  {\bf 44}, 1321 (2012)
  [arXiv:1104.2692 [hep-th]].

\bibitem{Zhang:2011uv} 
  Y.~l.~Zhang and M.~Sasaki,
  Int.\ J.\ Mod.\ Phys.\ D {\bf 21}, 1250006 (2012)
  [arXiv:1108.2112 [gr-qc]].


\bibitem{NL-Ref}
%
  L.~Parker and D.~J.~Toms,
  Phys.\ Rev.\ D {\bf 32}, 1409 (1985); 
%
  T.~Banks,
  Nucl.\ Phys.\ B {\bf 309}, 493 (1988);
%
  C.~Wetterich,
  Gen.\ Rel.\ Grav.\  {\bf 30}, 159 (1998)
  [gr-qc/9704052];
%
  A.~O.~Barvinsky,
  Phys.\ Lett.\ B {\bf 572}, 109 (2003)
  [hep-th/0304229];
  J.~Khoury,
  Phys.\ Rev.\ D {\bf 76}, 123513 (2007)
  [hep-th/0612052];
%
  L.~Joukovskaya,
  Phys.\ Rev.\  D {\bf 76}, 105007 (2007)
  [arXiv:0707.1545 [hep-th]];
%
  G.~Calcagni, M.~Montobbio and G.~Nardelli,
  Phys.\ Lett.\  B {\bf 662}, 285 (2008)
  [arXiv:0712.2237 [hep-th]];
%
  S.~Jhingan, S.~Nojiri, S.~D.~Odintsov, M.~Sami, I.~Thongkool and S.~Zerbini,
  Phys.\ Lett.\  B {\bf 663}, 424 (2008)
  [arXiv:0803.2613 [hep-th]];
%
  T.~Koivisto,
  Phys.\ Rev.\  D {\bf 77}, 123513 (2008)
  [arXiv:0803.3399 [gr-qc]]; 
  N.~A.~Koshelev,
  Grav.\ Cosmol.\  {\bf 15}, 220 (2009)
  [arXiv:0809.4927 [gr-qc]];
  M.~G.~Romania, N.~C.~Tsamis and R.~P.~Woodard,
  Lect.\ Notes Phys.\  {\bf 863}, 375 (2013)
  [arXiv:1204.6558 [gr-qc]]; 
%
  A.~O.~Barvinsky and Y.~V.~Gusev,
  Phys.\ Part.\ Nucl.\  {\bf 44}, 213 (2013)
  [arXiv:1209.3062 [hep-th]];
%
  S.~Deser and R.~P.~Woodard,
  JCAP {\bf 1311}, 036 (2013)
  [arXiv:1307.6639 [astro-ph.CO]]; 
%
\bibitem{Ivanov:2011vy}
  M.~M.~Ivanov and A.~V.~Toporensky,
  Grav.\ Cosmol.\  {\bf 18} (2012) 43
  [arXiv:1106.5179 [gr-qc]].



\bibitem{Boko:2016mwr}
  R.~D.~Boko, M.~J.~S.~Houndjo and J.~Tossa,
  Int.\ J.\ Mod.\ Phys.\ D {\bf 25} (2016) 1650098
  [arXiv:1605.03404 [gr-qc]].

\bibitem{Odintsov:2017icc}
  S.~D.~Odintsov, V.~K.~Oikonomou and P.~V.~Tretyakov,
  Phys.\ Rev.\ D {\bf 96} (2017) 044022
  [arXiv:1707.08661 [gr-qc]].

\bibitem{Odintsov:2015wwp}
  S.~D.~Odintsov and V.~K.~Oikonomou,
  Phys.\ Rev.\ D {\bf 93} (2016) 023517
  [arXiv:1511.04559 [gr-qc]].


\bibitem{Bohmer:2010re}
  C.~G.~Boehmer, T.~Harko and S.~V.~Sabau,
  Adv.\ Theor.\ Math.\ Phys.\  {\bf 16} (2012) 1145
  [arXiv:1010.5464 [math-ph]].

\bibitem{Goheer:2007wu}
  N.~Goheer, J.~A.~Leach and P.~K.~S.~Dunsby,
  Class.\ Quant.\ Grav.\  {\bf 24} (2007) 5689
  [arXiv:0710.0814 [gr-qc]].

\bibitem{Leon:2014yua}
  G.~Leon and E.~N.~Saridakis,
  JCAP {\bf 1504} (2015) 031
  [arXiv:1501.00488 [gr-qc]].

\bibitem{Leon:2010pu}
  G.~Leon and E.~N.~Saridakis,
  Class.\ Quant.\ Grav.\  {\bf 28} (2011) 065008
  [arXiv:1007.3956 [gr-qc]].

\bibitem{deSouza:2007zpn}
  J.~C.~C.~de Souza and V.~Faraoni,
  Class.\ Quant.\ Grav.\  {\bf 24} (2007) 3637
  [arXiv:0706.1223 [gr-qc]].

\bibitem{Giacomini:2017yuk}
  A.~Giacomini, S.~Jamal, G.~Leon, A.~Paliathanasis and J.~Saavedra,
  Phys.\ Rev.\ D {\bf 95} (2017) 124060
  [arXiv:1703.05860 [gr-qc]].

\bibitem{Kofinas:2014aka}
  G.~Kofinas, G.~Leon and E.~N.~Saridakis,
  Class.\ Quant.\ Grav.\  {\bf 31} (2014) 175011
  [arXiv:1404.7100 [gr-qc]].

\bibitem{Leon:2012mt}
  G.~Leon and E.~N.~Saridakis,
  JCAP {\bf 1303} (2013) 025
  [arXiv:1211.3088 [astro-ph.CO]].

\bibitem{Gonzalez:2006cj}
  T.~Gonzalez, G.~Leon and I.~Quiros,
  Class.\ Quant.\ Grav.\  {\bf 23} (2006) 3165
  [astro-ph/0702227].

\bibitem{Alho:2016gzi}
  A.~Alho, S.~Carloni and C.~Uggla,
  JCAP {\bf 1608} (2016) 064
  [arXiv:1607.05715 [gr-qc]].


\bibitem{Muller:2014qja}
  D.~Moller, V.~C.~de Andrade, C.~Maia, M.~J.~Rebouas and A.~F.~F.~Teixeira,
  Eur.\ Phys.\ J.\ C {\bf 75} (2015) 13
  [arXiv:1405.0768 [astro-ph.CO]].

\bibitem{Rippl:1995bg}
  S.~Rippl, H.~van Elst, R.~K.~Tavakol and D.~Taylor,
  Gen.\ Rel.\ Grav.\  {\bf 28} (1996) 193
  [gr-qc/9511010].

\bibitem{Ivanov:2011vy}
  M.~M.~Ivanov and A.~V.~Toporensky,
  Grav.\ Cosmol.\  {\bf 18} (2012) 43
  [arXiv:1106.5179 [gr-qc]].

\bibitem{Odintsov:2017icc}
  S.~D.~Odintsov, V.~K.~Oikonomou and P.~V.~Tretyakov,
  Phys.\ Rev.\ D {\bf 96} (2017) 044022
  [arXiv:1707.08661 [gr-qc]].

\bibitem{Odintsov:2015wwp}
  S.~D.~Odintsov and V.~K.~Oikonomou,
  Phys.\ Rev.\ D {\bf 93} (2016) 023517
  [arXiv:1511.04559 [gr-qc]].


\bibitem{Maggiore:2016gpx} 
  M.~Maggiore,
  Fundam.\ Theor.\ Phys.\  {\bf 187}, 221 (2017)
  [arXiv:1606.08784 [hep-th]].
\bibitem{Otalora:2016dxe} 
  G.~Otalora and E.~N.~Saridakis,
  Phys.\ Rev.\ D {\bf 94}, no. 8, 084021 (2016)
  doi:10.1103/PhysRevD.94.084021
  [arXiv:1605.04599 [gr-qc]].
\bibitem{Bahamonde:2017bps} 
  S.~Bahamonde, S.~Capozziello, M.~Faizal and R.~C.~Nunes,
  Eur.\ Phys.\ J.\ C {\bf 77}, 628 (2017)
  [arXiv:1709.02692 [gr-qc]]. 
\bibitem{Channuie:2017txg} 
  P.~Channuie and D.~Momeni,
  Nucl.\ Phys.\ B {\bf 935}, 256 (2018)
  doi:10.1016/j.nuclphysb.2018.08.016
  [arXiv:1712.07927 [gr-qc]].
  
  
\bibitem{Hayashi}
K.~Hayashi and T.~Shirafuji,
  Phys.\ Rev.\ D {\bf 19}, 3524 (1979)
  Addendum: [Phys.\ Rev.\ D {\bf 24}, 3312 (1982)]
\bibitem{Bahamonde:2017sdo} 
  S.~Bahamonde, S.~Capozziello and K.~F.~Dialektopoulos,
  Eur.\ Phys.\ J.\ C {\bf 77}, 722 (2017)
  [arXiv:1708.06310 [gr-qc]].

\bibitem{Jamil:2012nma} 
  M.~Jamil, D.~Momeni and R.~Myrzakulov,
  Eur.\ Phys.\ J.\ C {\bf 72}, 1959 (2012)
  [arXiv:1202.4926 [physics.gen-ph]].

\bibitem{Jamil:2012vb} 
  M.~Jamil, D.~Momeni and R.~Myrzakulov,
  Eur.\ Phys.\ J.\ C {\bf 72}, 2075 (2012)
  [arXiv:1208.0025 [gr-qc]].

\bibitem{Awad:2017yod} 
  A.~Awad, W.~El Hanafy, G.~G.~L.~Nashed and E.~N.~Saridakis,
  arXiv:1710.10194 [gr-qc].

\bibitem{Jamil:2012yz} 
  M.~Jamil, K.~Yesmakhanova, D.~Momeni and R.~Myrzakulov,
  Central Eur.\ J.\ Phys.\  {\bf 10}, 1065 (2012)
  [arXiv:1207.2735 [gr-qc]].

\bibitem{arXiv:1012.4879}
Pan Yu, Li Li, Cao Shuo, Pan Na-na, Zhang Yi, Hu Zi-xuan, "Testing the interaction between dark energy and dark matter with H(z) data",  Chinese Astronomy and Astrophysics, Volume 40, Issue 2, 2016, Pages 176-185, ISSN 0275-1062, https://doi.org/10.1016/j.chinastron.2016.05.007.

\bibitem{interaction}
%
  N.~Cruz, S.~Lepe and F.~Pena,
  Phys.\ Lett.\ B {\bf 663}, 338 (2008)
  [arXiv:0804.3777 [hep-ph]];
%
  N.~Cruz, S.~Lepe and F.~Pena,
  Phys.\ Lett.\ B {\bf 699}, 135 (2011);\\ 
%
  M.~Jamil, E.~N.~Saridakis and M.~R.~Setare,
  Phys.\ Rev.\ D {\bf 81}, 023007 (2010)
  [arXiv:0910.0822 [hep-th]];
%
  M.~Jamil, D.~Momeni and M.~A.~Rashid,
  Eur.\ Phys.\ J.\ C {\bf 71}, 1711 (2011)
  [arXiv:1107.1558 [physics.gen-ph]]; 
%
  S.~Chen and J.~Jing,
  Class.\ Quant.\ Grav.\  {\bf 26}, 155006 (2009)
  [arXiv:0903.0120 [gr-qc]].
%

\bibitem{He:2010im} 
  J.~H.~He, B.~Wang and E.~Abdalla,
  Phys.\ Rev.\ D {\bf 83}, 063515 (2011)
  [arXiv:1012.3904 [astro-ph.CO]].

\bibitem{dynamics}
A. A. Coley, Dynamical systems and cosmology, vol. 291. Kluwer, Dordrecht, Netherlands, 2003, 10.1007/978-94-017-0327-7; 
G. Leon and C. R. Fadragas, Cosmological dynamical systems. LAP Lambert Academic Publishing, 2012; 
  C.~G.~Boehmer and N.~Chan,
  ``Dynamical systems in cosmology,'' 
  arXiv:1409.5585 [gr-qc];
J.~Dutta, W.~Khyllep, E.~N.~Saridakis, N.~Tamanini and S.~Vagnozzi,
arXiv:1711.07290 [gr-qc].


\bibitem{sn}
  M.~Betoule {\it et al.} [SDSS Collaboration],
  Astron.\ Astrophys.\  {\bf 568}, A22 (2014)
  [arXiv:1401.4064 [astro-ph.CO]].

\bibitem{bao1} 
F. Beutler, C. Blake, M. Colless, D. H. Jones, L. Staveley-Smith, L. Campbell, Q. Parker, W. Saunders and F. Watson,
  Mon.\ Not.\ Roy.\ Astron.\ Soc.\  {\bf 416}, 3017 (2011)
  [arXiv:1106.3366 [astro-ph.CO]].

\bibitem{bao2} 
  A.~J.~Ross, L.~Samushia, C.~Howlett, W.~J.~Percival, A.~Burden and M.~Manera,
  Mon.\ Not.\ Roy.\ Astron.\ Soc.\  {\bf 449} 835 (2015)
  [arXiv:1409.3242 [astro-ph.CO]].

\bibitem{bao3}
  L.~Anderson {\it et al.} [BOSS Collaboration],
  Mon.\ Not.\ Roy.\ Astron.\ Soc.\  {\bf 441} 24 (2014)
  [arXiv:1312.4877 [astro-ph.CO]].

\bibitem{bao4}
  A.~Font-Ribera {\it et al.} [BOSS Collaboration],
  JCAP {\bf 1405}, 027 (2014)
  [arXiv:1311.1767 [astro-ph.CO]].

\bibitem{baotot} 
  R.~C.~Nunes, S.~Pan, E.~N.~Saridakis and E.~M.~C.~Abreu,
  JCAP {\bf 1701} 005 (2017)
  [arXiv:1610.07518 [astro-ph.CO]].

\bibitem{cc}
  M.~Moresco, R.~Jimenez, L.~Verde, A.~Cimatti, L.~Pozzetti, C.~Maraston and D.~Thomas,
  JCAP {\bf 1612} 039 (2016)
  [arXiv:1604.00183 [astro-ph.CO]].

\bibitem{Capozziello:2011hj} 
  S.~Capozziello, V.~F.~Cardone, H.~Farajollahi and A.~Ravanpak,
  Phys.\ Rev.\ D {\bf 84}, 043527 (2011)
  [arXiv:1108.2789 [astro-ph.CO]].

\bibitem{Wu:2011xa}
  P.~Wu and H.~Yu,
  Phys.\ Lett.\  B {\bf 703}, 223 (2011)
  [arXiv:1108.5908 [gr-qc]].

\end{thebibliography}
\end{document}